# Control of fluctuation in ECR plasma by applying multipole magnetic fields

Hayato Tsuchiya[a], Mayuko Koga, Masayuki Fukao, and Yoshinobu Kawai

[a]*Interdisciplinary Graduate School of Engineering Sciences,*
*Kyushu University, Kasuga, Fukuoka 816-8580, Japan*

The control of fluctuations generated in an electron cyclotron resonance (ECR) plasma is attempted by changing microwave powers. It is found that when a cage is introduced in the vacuum chamber, the fluctuations change from turbulent state to chaotic state in the high microwave power region while it changes from chaotic state to periodic state in the low microwave power region. This result suggests that the geometry of the cage affects the state of the instability. Moreover it is found that the superposition of multicusped fields created by small permanent magnets reduces the chaotic dimension of the instability. When pulsed multipole magnetic fields are also added to external magnetic fields, the fluctuations are suppressed.

## 1. Introduction

There is an increasing interest in chaos and turbulence generated in plasmas [1]. Chaos and turbulence are often troublesome phenomena which may have harmful consequence. Such systems reach turbulent state via chaos. Therefore, the investigation of controlling chaos will contribute to an understanding of turbulence in fusion-oriented plasmas. Chaos in high temperature plasmas will evolve into fully developed turbulence and lead to the anomalous transport. Suppression of turbulence is necessary for improving the confinement.

There are some reports about the controlling of instabilities in plasma [2-5]. Especially, chaos control is one of the most powerful methods because it does not need high power to control [6-7]. However, chaos control is only effective when the system is in chaos state. Therefore, the transition of the system from turbulent state to chaotic state is important for chaos control. Matsukuma et al. found that the ion-ion instability excited in double plasma was changed from turbulent state to chaos state by applying a mesh grid as a boundary condition [8]. Itoh et al. examined the role of a boundary condition by simulation and reported that a certain boundary condition upon the system reduced the number of degrees of freedom and as the result the state of system changed [9]. Thus, it is considered that the boundary condition has great influence on the state of instability.

Recently, we found that the density fluctuation in electron cyclotron resonance

(ECR) plasma [10-12] was excited by flute instability and the instability was suppressed by superimposing a permanent magnet cage [12]. Moreover, it was found that the superimposing of the multicusped fields formed by the permanent magnet cage changed the behavior of the fluctuation from turbulent state to chaos state [1], suggesting that the permanent magnet cage may play a role of a boundary condition in our preliminary experiment. In the previous experiment, however, it was not clear whether or not the change of the instability was caused by the multicusped fields formed by the permanent magnet cage. Furthermore, the permanent magnet cage may change the geometry of the experimental system, that is, there is a possibility that the change of the geometry caused the chaos state of the instability.

Since the permanent magnet cage can not vary the intensity of magnetic fields, the coils which generate the multipole fields is put into place around the vacuum chamber.

## 2. Experimental setup

Figure.1 shows a schematic diagram of the experimental setup of ECR plasma. The vacuum chamber is made of stainless-steal whose size is 400mm in inner diameter and 1200mm in length. Eight magnetic coils are used to form the magnetic mirror field. The chamber is evacuated using a rotary pump and a turbomolecular pump to abase pressure of less than $3 \times 10^{-6}$ Torr. A magnetron generates the microwaves (2.45GHz, max power 1kW), which can modulate the power by external signal. The microwaves are launched as the circular $TE_{11}$ mode into the chamber through the tapered waveguide and the quartz window. The used gas is Argon. The plasma parameters are measured with a cylindrical Langmuir probe (1 mm in diameter and 1mm in length). Experimental gas pressure is around 0.5 mTorr. The ECR plasma generated by the microwaves of 500W has following features, the electron density $n_e \sim 10^{10}[cm^{-3}]$ and the electron temperature $T_e \sim 4[eV]$.

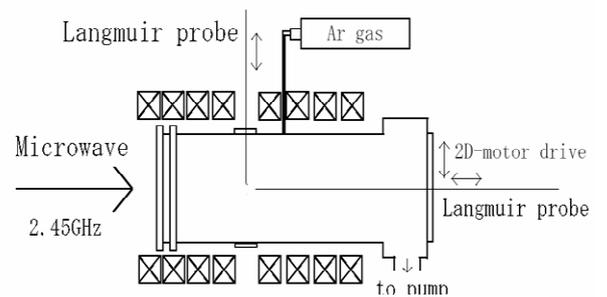

Fig.1 Schematic diagram of the experimental setup of ECR plasma

The magnetic cage consisted of magnetic pillars is shown in Fig. 2. The magnetic pillars are made from stainless-covered permanent magnets, whose dimensions are 24 mm square and 561 mm in length. The pillars are aligned to form the multicusped magnetic field. The dimensions of the cage are 290 mm in diameter and 500 mm in length. The cage without magnetic field (dummy magnetic cage) is obtained by replacing magnetic pillars with stainless pillars that did not contain permanent magnets.

Fig. 3 is a schematic diagram of the coil which generates the multipole magnetic fields located around the vacuum chamber. Pulsed electric current is flown by the eight-condensers of capacitances of 37.6 mF. The maximum charge voltage is 300 V. The intensity of the multipole fields is approximately 150 gauss at measurement point against the background magnetic field 650 gauss.

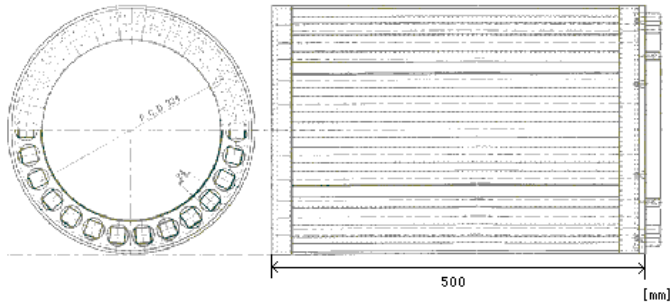
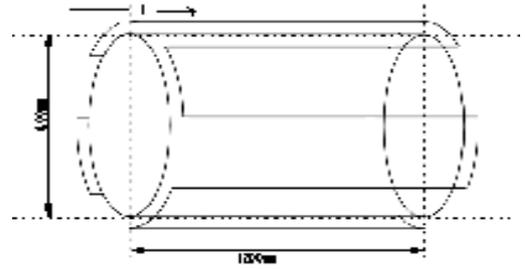

Fig.2 Schematic diagram of the magnet cage          Fig.3 Schematic diagram of the multipole magnetic coils

## 3. Experimental Results
### 3.1 Applying the multicusped fields

Figure 4 (a) shows time series data of the floating potential fluctuation without the cage. The frequency of the fluctuation was about 4.5 kHz. It was confirmed that the observed fluctuation had characteristics of the flute instability [13], i.e., the phase difference between density fluctuation and potential fluctuation was out of phase at the same point, the fluctuation grew at the point where the density profile had the gradient, and the phase of the fluctuation was constant along the direction parallel to the magnetic field. As shown in Fig. 4 (a), time series data shows relatively periodic behavior in the low microwave power region, 150 W, and becomes chaotic behavior as the incident microwave power is increased.

Then, we performed chaos analysis from time series data shown in Fig. 4 (a) in order to evaluate whether the system was in chaos state or not. A correlation dimension was calculated by using the method of Grassberger and Procaccia [14]. If the correlation dimension saturates to an integer with increasing the embedding dimension, the system is periodic and if it saturates to a fraction, the system is chaotic. If it does not saturate, the system is turbulent. The calculated result is shown in Fig. 4 (b). The correlation dimension saturates to non-integer 3.5 for 150 W, wheile it does not saturate for 450 W. Therefore, it is found that the instability changed from chaos state to turbulent state as the incident microwave power was increased.

Then, the effect of multicusped fields on the fluctuation is confirmed by inserting the magnetic cage into the chamber. Figure 5 shows (a) time series data and (b) the calculated correlation dimension of the floating potential fluctuation with the multicusped magnetic fields. Comparing Fig. 5 (a) with Fig. 4 (a), it is clearly seen that the behavior of the fluctuation changes more periodic for 150 W. This change corresponds to the result of calculated correlation dimensions. As shown in Fig. 5 (b), the correlation dimension in the presence of the magnetic cage saturates to integer, 3, that is, the system is in periodic state. On the other hand, the correlation dimension saturates to non-integer 4.2 even if the incident microwave power is increased up to 450 W. These results suggest the stabilization of the instability was achieved by using the magnetic cage. Moreover, comparing to Fig.4 (a), note that the amplitude of the fluctuation is suppressed when the magnetic cage is superimposed for 150 W, while it does not change very much for 450 W. From this result, it is considered that the suppression of the instability by the magnetic cage is effective in the low microwave power region.

In order to examine the effect of magnetic fields, the dummy magnetic cage is replaced with the magnetic cage. This cage consisted of stainless pillars whose size is the same as that of the stainless-covered magnet pillars. The results are shown in Fig. 6. The calculated correlation dimension shows a non-integer 5.5 for 450 W, while it changes to an integer 5 for 150 W. This means that the state of fluctuation changed from chaotic state to periodic state by adding the dummy magnetic cage. Thus, it is considered that the state of the fluctuation changes by the geometry of the cage rather than the magnetic field of the cage. The change of the plasma edge condition by adding the cage is considered to be the main reason for the change of the state of the instability. Moreover, it is reported that the reduction of the system size associates with the positive Lyapunov exponent [15]. Therefore, the addition of the cage is considered to be effective to control the chaotic state of the instability. However, the correlation dimension is higher compared with the case of using the magnetic cage. This result indicates that the magnetic fields of the cage reduce the chaotic dimension of the instability. It is considered that the magnetic fields restrict the particle motion and reduce the number of degrees of freedom. Therefore, the reduction of the number of degrees of freedom may result in the reduction of the chaotic dimension. Moreover, the magnetic field may affect the instability directly, because the flute instability is caused by the bad curvature of the magnetic field.

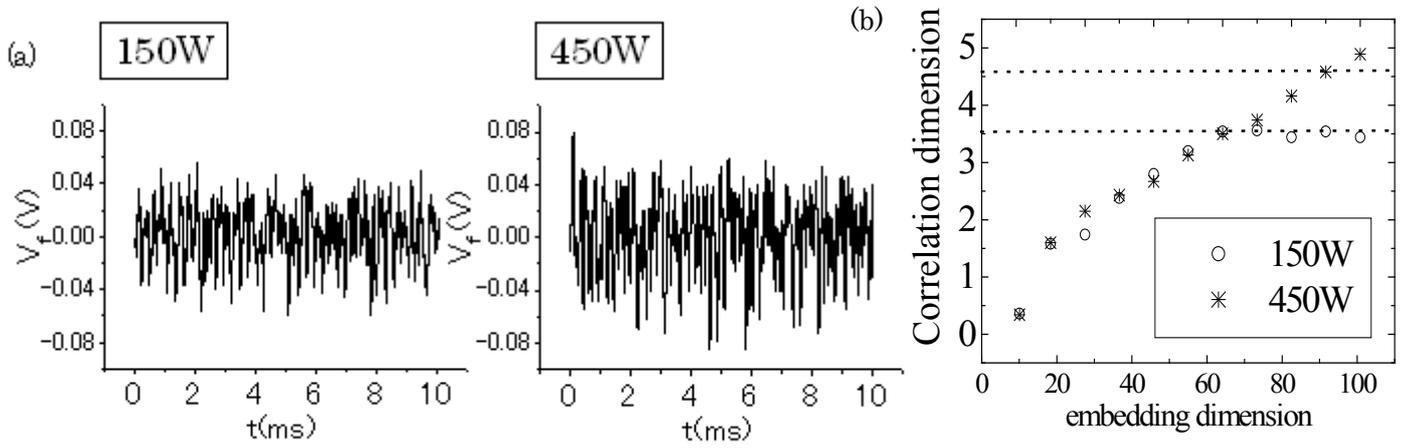

Figure 4 (a) Time series data and (b) calculated correlation dimension of the fluctuation without the cage

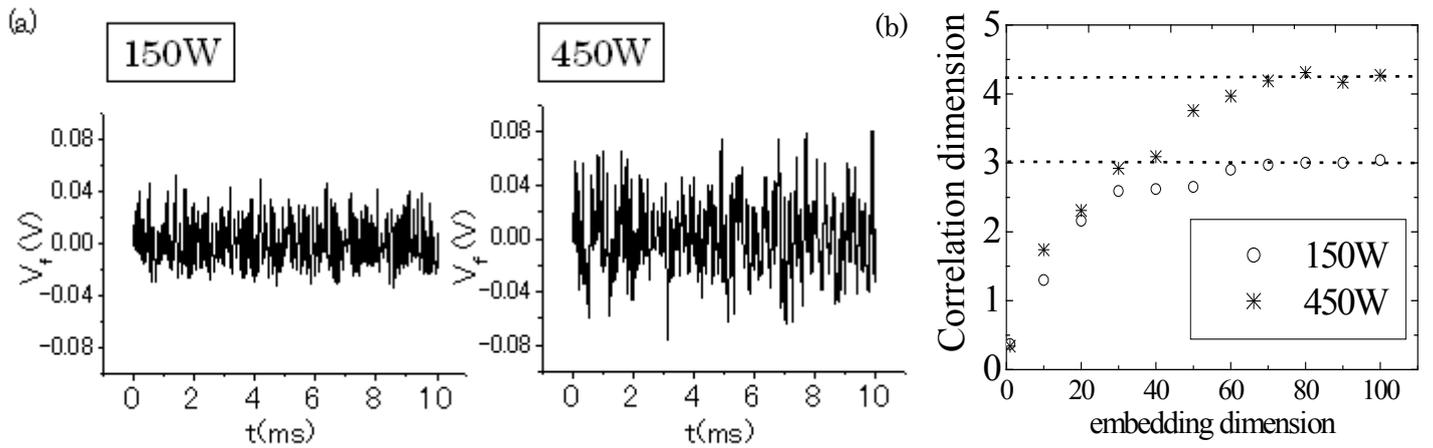

Figure 5 (a) Time series data and (b) calculated correlation dimension of the fluctuation with the magnetic cage

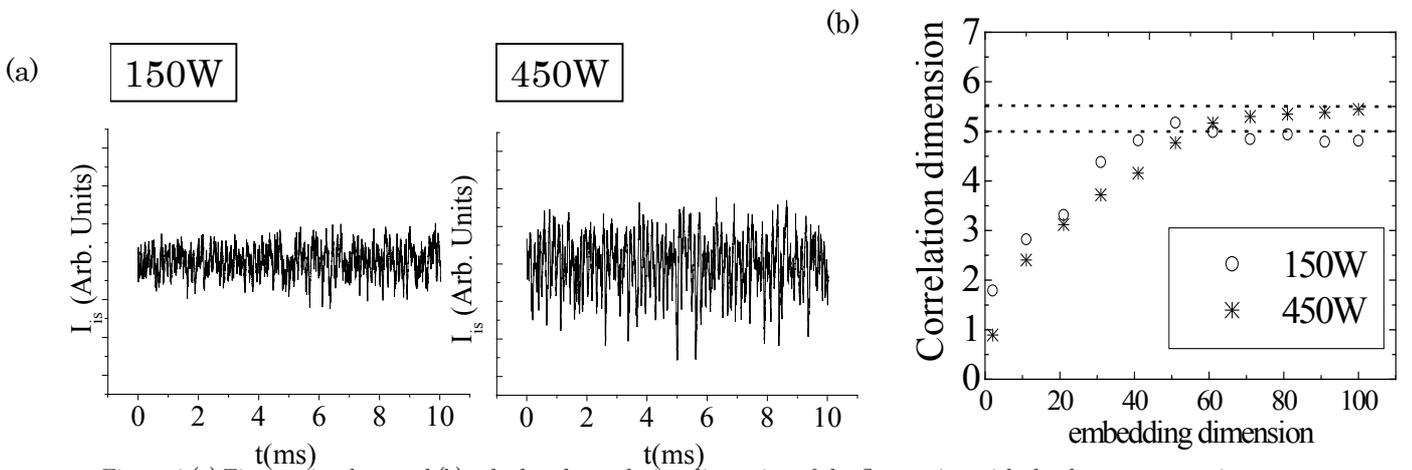

Figure 6 (a) Time series data and (b) calculated correlation dimension of the fluctuation with the dummy magnetic cage

### 3.2 Applying multipole magnetic field

To control the fluctuations multipole magnetic fields are added into the mirror magnetic fields. The multipole magnetic fields are created by pulsed coil current with a coil set up around the chamber. The intensity of the multipole fields is approximately 150 gauss at measurement point against the background magnetic field 650 gauss. Fig.8 shows the time series data of $I_{is}$ when the multipole magnetic fields are applied.

The upper figure shows the fluctuation of the ion saturation current, the lower one shows the intensity of the multipole magnetic fields. From this figure it is clear to suppress the amplitude of the fluctuations with the adding multipole magnetic fields. The amplitude of the fluctuations is half as high as that of the fluctuation without pulsed multipole magnetic fields. The suppression is not caused by the flatness of the radial profile of the plasma density. On the contrary it was observed that the density gradient becomes a little steep. It is concludeed that the improvement of the magnetic configuration by the pulsed multipole magnetic fields has resulted in the suppression of the fluctuations.

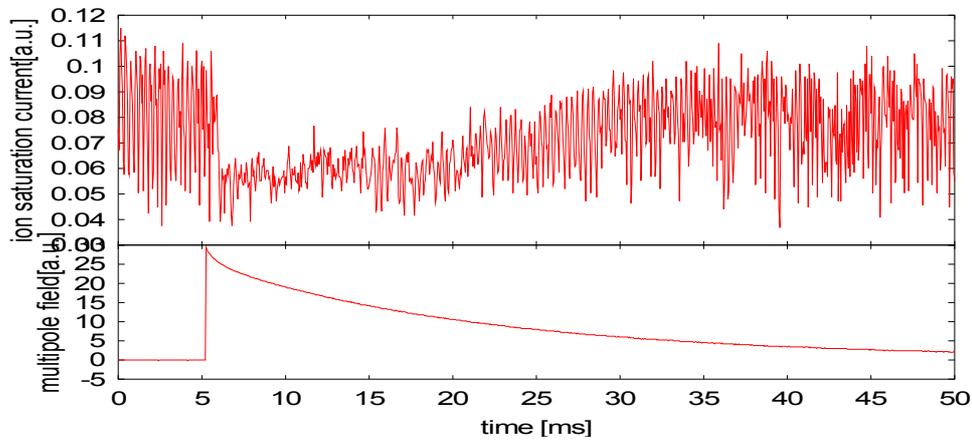

Fig.7 suppression of the fluctuation cased by multipole magnetic fields; lower graph shows the intensity of the multipole magnetic field. Upper graph shows the fluctuations of the ion saturation current.

## 4 Conclusions

We observed the fluctuations caused by the flute instability in an ECR plasma. The behavior of the instability changed from chaotic state to turbulent state, as incident microwave power was increased. It was found that the magnetic cage stabilized the instability from chaotic state to periodic state in the low microwave power region, from turbulent state to chaotic state in the high microwave power region. It was also found that the dummy magnetic cage also stabilized the instability, although the chaotic dimension was high compared with the case of using magnetic cage. Therefore, it was considered that the change of the plasma edge condition due to the geometry of the cage was the main reason for the change of the state of the instability. And with regard to the addition of the multicusped fields, it reduced the number of degrees of freedom of the system and reduced the chaotic dimension of the instability.

The fluctuations were suppressed with the use of the multipole magnetic fields,

which was caused by improvement of the minimum *B* configuration. This experimental result indicates that the magnetic configuration is also one of the dominant factors for controlling chaos. Furthermore, there may be some methods to control chaos using electric fields. Very recently we have started the experiments on chaos control by modulating the microwave power.